\documentclass[9pt,conference]{IEEEtran}
\usepackage[preprint]{dcase2026}

\usepackage{bm} 
\usepackage{xcolor}
\usepackage{multirow}

\usepackage{dcase2026,amsmath,graphicx,url,times,booktabs, tabularx}

\title{Drowning in Noise: On Signal Saliency in Self-Supervised Learning for Ocean Acoustics}

\name{Shikhar Bharadwaj$^{1}$,
      Samuele Cornell$^{1}$,
      Shinji Watanabe$^{1}$,
      Malcolm Slaney$^{2}$}
\address{$^{1}$Carnegie Mellon University, USA \;
$^{2}$Stanford, USA\\
}

\begin{document}

\maketitle
\begin{abstract}
Self-supervised learning (SSL) for audio typically benefits from larger pre-training corpora, but this relationship assumes that training data carries meaningful learning signal.
In ocean acoustics, this assumption is often violated: passive recordings are dominated by waves, wind, and silence, with biological vocalizations occupying a small fraction of the total duration.
We show that this \emph{overrepresentation} of noise causes SSL models to allocate representational capacity to acoustically trivial structure, degrading downstream performance.
Using BEATs as a testbed, we demonstrate that pre-training on 30 hours of \emph{salient} recordings (segments containing animal vocalizations) outperforms 1{,}240 hours of unfiltered data on a 31-class marine mammal classification benchmark.
Under linear probing, the manually curated salient set achieves markedly better accuracy than the full corpus, despite using $41\times$ less data.
Automatic saliency-based filtering improves over unfiltered pre-training, but still falls well short of the manually curated salient set.
Our findings suggest that, in domains where informative content is temporally sparse, data curation matters more than data scale for SSL pre-training.
\end{abstract}


\section{Introduction}

The success of large-scale self-supervised learning (SSL) in speech~\cite{hubert, wav2vec} and general audio~\cite{audiomae, beats} has prompted interest in applying foundation models to specialized domains, including ecological monitoring and marine biology~\cite{hagiwara2023aves, deng2025gpmbT}.
The prevailing intuition, borrowed from the scaling-law literature~\cite{kaplan2020scaling}, is that more pre-training data yield stronger representations~\cite{openbeats, dasheng}.
However, for passive monitoring applications such as ocean acoustics, this intuition can fail: in long recordings, meaningful signals occupy only a tiny fraction of the total duration.
A 40-second clip from the Watkins Marine Mammal Sound Database~\cite{sayigh2016watkins}, for instance, may contain only three distinct animal vocalizations, the rest being broadband ocean noise.

We show that, when an SSL model is trained on such data using conventional methods, it devotes the majority of its representational capacity to modeling noise structure rather than the saliency of the signals of interest.
We call this phenomenon \emph{noise overrepresentation} and provide the first study on its impact on SSL for ocean acoustics.
This connects to a broader observation in the scaling literature: beyond a point, indiscriminately adding data yields diminishing---or negative---returns, and careful pruning of low-signal examples can match or beat scale~\cite{sorscher2022pruning, oquab2024dinov2}.
We stress that this is a distinct axis from \emph{domain match}: recent bioacoustic studies find that out-of-domain data at scale can transfer surprisingly well, with encoders pre-trained on large bird corpora rivaling or exceeding in-domain training on marine tasks~\cite{tropicalreef_bird_transfer_learning, van2025perch}. Our question is orthogonal to this axis. Holding the source corpus fixed, we ask how the \emph{temporal density} of informative events---not the domain they are drawn from---shapes the learned representation.

Two lines of work are closest to ours, both in bird bioacoustics. AVES-bio~\cite{hagiwara2023aves} filters training data using ground-truth labels, showing that curating for the downstream task yields higher-quality representations; and Moummad \cite{birdfilterssl} uses PANN~\cite{kong2020panns} activations as a pseudo-label to select segments for SSL pre-training on bird calls~\cite{birdclef}. Neither examines how the \emph{noise} introduced by imperfect, automatic filtering propagates into the learned representation, nor addresses the ocean domain---the two gaps we study. We explicitly ablate salient event detection methods for filtering against a label-driven oracle in the ocean setting.
More generally, the salient event detection methods developed for bird bioacoustics have not been tested on ocean sounds, where broadband noise makes the problem substantially harder.
In ocean acoustics, the general trend is to pre-train on AudioSet~\cite{gemmeke2017audioset}, which already consists of high-density clips~\cite{oceanssl2023}. 
Meanwhile, data from large-scale passive ocean acoustic monitoring systems remain untapped, one major hurdle being the sparsity of informative signals they contain.
Our work explores a benchmark for ocean salient event detection and provides an exploratory analysis of the impact of treating saliency as a first-class citizen in ocean foundation modeling.

To summarize, we make the following contributions:
\begin{enumerate}
\item We provide the first controlled study of noise overrepresentation in SSL pre-training for ocean acoustics, demonstrating that a $41\times$ reduction in training data, by retaining only salient events, yields a \textbf{29-point improvement} in linear probing accuracy on marine mammal classification.
\item We compare salient event detection filtering methods for ocean acoustics. Results show that most methods perform poorly because of the ocean's characteristic broadband noise. Ablations quantify these gaps and motivate the need for better signal saliency detection tailored to ocean acoustics.
\item We introduce the concept of \emph{signal saliency} as a primary criterion for pre-training data curation in domain-specific foundation models, arguing that it is orthogonal to, and often more impactful than, dataset scale.
\end{enumerate}

Our findings have immediate practical implications: building acoustic foundation models does not require collecting vast quantities of passive recordings.
It requires collecting recordings that contain sounds worth learning from.
More broadly, this demonstrates that wherever the signal is sparse against a dominant background, such as in most continuous passive monitoring applications, curation should be a first-class step in pre-training, since the density of signal, not the volume of data, is the bottleneck.

\section{Background}

\subsection{Audio Representation Learning}
Modern SSL frameworks for audio generally operate by masking portions of the input and training an encoder to reconstruct~\cite{audiomae} or predict discrete targets for those masked regions~\cite{beats, hubert, hagiwara2023aves, deng2025gpmbT}.
Among these, BEATs~\cite{beats} patches the mel spectrogram into an 8$\times$8 grid, feeds the patches into a Vision Transformer (ViT)~\cite{dosovitskiy2021vit}, and iterates between training the SSL encoder (via masked token prediction) and an acoustic tokenizer (via knowledge distillation from the encoder).
It is a strong, widely-adopted general-audio SSL model: for example it was the standard pre-trained embedding in the DCASE Task~4 sound event detection challenge~\cite{cornell2024dcase} and serves as the non-speech audio encoder in audio-language models such as SALMONN~\cite{tang2024salmonn}. Its open reproduction, OpenBEATs~\cite{openbeats, openbeatsicme}, extends it to other domains, including bioacoustics.
Because of its broad adoption and strong performance, in this work we chose BEATs as our testbed to study the effect of training data composition.

\subsection{Salient Event Detection for Representation Learning}
\label{sec:sed}

Ocean passive acoustic recordings differ from the corpora on which general audio SSL models are trained in one critical respect: the \emph{duty cycle} of informative content is extremely low. 
AudioSet clips are typically centered on an identifiable sound event while passive ocean recordings continuously capture the ambient environment.
Biological vocalizations, the signals that matter for most downstream tasks, may occupy less than 5\% of total recording duration in some datasets~\cite{sayigh2016watkins}, while silence, wind, water and static noise dominate.

This creates a mismatch that is not merely a domain shift in the conventional sense.
The model does not simply encounter unfamiliar sounds; it is trained to model sounds that are irrelevant to the tasks it will be evaluated on. The question we investigate is: \emph{how severely does this hurt, and what does the hurt look like mechanistically?}

The closest domain to ocean bioacoustics is perhaps bird bioacoustics, where self-supervised models are pre-trained on filtered data.
For instance, BirdSet~\cite{birdset_iclr} derives its event list with the \texttt{bambird} package~\cite{bambird}, which is built on \texttt{scikit-maad}'s region-of-interest segmentation~\cite{ulloa2021scikit, hamer2023birb}.
Models such as Bird-MAE~\cite{birdmae} then pre-train on this event selection.

However, these detectors were not designed for ocean acoustics, which differs from bird bioacoustics in both duty cycle and the kind of background noise. 
Ocean acoustics span a far wider band: marine mammal vocalizations range from infrasonic baleen calls to ultrasonic odontocete clicks above $100$\,kHz, whereas birdsong is largely confined to $1$--$10$\,kHz. 
Passive ocean recordings are also sparse in salient events, whereas bird soundscapes are often continuously active. 
Ultimately, the learned representation effectiveness is bounded by the quality of dataset filtering.
\textit{Whether detectors developed and deployed on birdsong remain useful under water} is therefore an open question, which we answer by benchmarking directly in this work.

\subsection{BEATs Pre-training Procedure}
\label{sec:beats}

BEATs~\cite{beats} is an iterative SSL framework that alternates between two stages: an encoder trained by masked token prediction, and an acoustic tokenizer trained by knowledge distillation from the encoder, repeated for up to three iterations so that tokenizer and encoder co-adapt to the training corpus.
In our experiments we \emph{do not re-run tokenizer training on ocean data}; we hold the tokenizer fixed so that the encoder is the only component that adapts to the pre-training corpus, and vary the tokenizer as a separate, controlled factor (Sec.~\ref{sec:rq1}). 
We describe the encoder objective below, and the tokenizer only to the extent needed to interpret our results.

\noindent\textbf{Encoder training:} The BEATs encoder follows a masked language modeling (MLM) paradigm applied to audio.
Given an input sequence of $d$-dimensional mel spectrogram patches of length $N$, $\mathbf{X} = \{\mathbf{x}_n \in \mathbb{R}^d\}_{n=1}^N$, a random subset of patches is masked, producing a sequence of unmasked patches $\tilde{\mathbf{X}}$.
The encoder produces $h$-dimensional embeddings $f^e(\tilde{\mathbf{X}}) = \{\mathbf{o}_n \in \mathbb{R}^{h}\}_{n=1}^N$, which are passed through a label-predictor module $f^e_\text{pred}$.
The composition $f^e_\text{pred} \circ f^e$ is trained to predict the masked tokens by minimizing the cross-entropy between the predicted and ground-truth token indices, given the unmasked patches:
\begin{equation}
\mathcal{L}_\text{MLM} = - \sum_{n \in M} \log p(z_n \mid \tilde{\mathbf{X}}),
\label{eq:mlm_loss}
\end{equation}
where $M$ is the set of masked positions and $z_n$ is the ground-truth token for position $n$ produced by the fixed tokenizer. In the first BEATs iteration the tokens come from a random-projection quantizer~\cite{bestrq}; from the second iteration onward, a learned tokenizer is used.

\noindent\textbf{Acoustic tokenizer:} The BEATs tokenizer is trained separately, in a teacher-student fashion, where the teacher is the encoder from the previous iteration. 
It maps each patch embedding to the nearest entry of a learned codebook and is optimized with a knowledge-distillation loss that aligns the tokenizer with the teacher, in the style of the vector-quantized variational autoencoder (VQ-VAE)~\cite{van2017neural}. 
Because this distillation makes the tokenizer encode corpus-specific semantics, a tokenizer co-adapted to a noise-dominated corpus and one co-adapted to a curated corpus would define different target vocabularies; holding the tokenizer fixed avoids this and is what makes the two conditions in Section~\ref{sec:rq1} directly comparable.
The token sequence a given segment induces is then a \emph{property of the segment}, and we study the token sequence in the failure-mode analysis in Sec.~\ref{sec:rq4}.

\subsection{The Overrepresentation Problem}

Let $\mathcal{D} = \{x_i\}_{i=1}^{N}$ be a pre-training corpus of audio segments.
In general-domain SSL, it is reasonable to assume that all segments carry comparable learning signal---every clip contains some sound event.
In ocean acoustics, the corpus decomposes as $\mathcal{D} = \mathcal{D}_s \cup \mathcal{D}_n$, where $\mathcal{D}_s$ contains segments with salient biological vocalizations and $\mathcal{D}_n$ contains noise-dominated segments, with $|\mathcal{D}_n| \gg |\mathcal{D}_s|$.
When an SSL model minimizes its training objective over $\mathcal{D}$, it implicitly allocates representational capacity proportional to the frequency of the patterns it must predict.
If $|\mathcal{D}_n|/|\mathcal{D}| \approx 0.95$, the encoder will spend the majority of its capacity learning to represent noise, leaving comparatively little for biological signals of interest.
We explore this problem in Sec.~\ref{sec:rq4}.

\section{Experiments}

We investigate four research questions (RQ): the effect of saliency filtering on learned representations (RQ1, Sec.~\ref{sec:rq1}); how well existing salient event detectors transfer to ocean audio (RQ2, Sec.~\ref{sec:rq2}); whether salient-event based filtering can replace human annotation (RQ3, Sec.~\ref{sec:rq3}); and the mechanism by which non-salient audio consumes model capacity (RQ4, Sec.~\ref{sec:rq4}).

\begin{table*}[t]
\caption{Marine mammal classification accuracy (\%) on the 31-class Watkins test set (BEANS)~\cite{hagiwara2023beans}, under linear probing (frozen encoder) and full fine-tuning. Setting~0 is the no-pre-training baseline; Settings~1--4 and 5--8 pair each corpus with the Random Codebook and the BEATs-iter3 Tokenizer. Best per column within each tokenizer block in \textbf{bold}. 
The 30\,hr Cut Tapes beat the 1{,}240\,hr Master Tapes by $\sim$29 points under linear probing with $41\times$ less data. 
Please see Sec.~\ref{sec:rq1} and Sec.~\ref{sec:rq3}.}\label{tab:results}
\centering
\small
\begin{tabular*}{\textwidth}{
    @{\extracolsep{\fill}}
    c l c l c c
    @{}
}
\toprule
\textbf{Setting}
& \textbf{Pre-training Data}
& \textbf{Volume (hr)}
& \textbf{Tokenizer}
& \textbf{Linear Probing}
& \textbf{Full Fine-tuning} \\
\midrule
0 & -                  & 0    & -               & 33.6 & 77.9 \\
\midrule
1 & AudioSet           & 5388 & Random Codebook & 65.5 & \textbf{93.2} \\
2 & Watkins (Master Tapes)       & 1240 & Random Codebook & 43.1 & 79.4 \\
3 & Watkins (Saliency Filtered)  & 613  & Random Codebook & 47.5 & 90.6 \\
4 & Watkins (Cut Tapes)          & 30   & Random Codebook & \textbf{72.0} & 91.2 \\
\midrule
5 & AudioSet  & 5388 & BEATs Tokenizer  & 65.8 & \textbf{92.3} \\
6 & Watkins (Master Tapes)       & 1240 & BEATs Tokenizer  & 42.8 & 87.0 \\
7 & Watkins (Saliency Filtered)  & 613  & BEATs Tokenizer  & 63.1 & 89.7 \\
8 & Watkins (Cut Tapes)          & 30   & BEATs Tokenizer  & \textbf{73.7} & 92.0 \\
\bottomrule
\end{tabular*}
\end{table*}

\subsection{RQ1: Impact of Dataset Filtering on Representation Learning}
\label{sec:rq1}

\noindent \textbf{Setup:} We construct two training datasets using audio from the Watkins Marine Mammal Sound Database~\cite{sayigh2016watkins}: 1) \textbf{Master Tapes:} Unfiltered recordings containing the full signal with biological vocalizations, ambient ocean noise, wind, and silence over a total duration of \textbf{1{,}240 hours}. 2) \textbf{Cut Tapes:} Recordings that are operator-selected excerpts of Master Tapes that retain only segments containing confirmed biological vocalizations with a total duration of \textbf{30 hours}.

Using these datasets, we pre-train two models following the BEATs framework~\cite{beats, openbeats} to study how data composition affects learned representations.
To disentangle the effects of data saliency from tokenizer quality, we cross each pre-training dataset with two tokenizer choices: 1)
\textbf{Random Codebook:} Token targets are assigned using a random projection, with no learned structure.
2) \textbf{BEATs Tokenizer:} Token targets are produced by the BEATs iter3 tokenizer trained on AudioSet.
We also include the BEATs model pre-trained on AudioSet~\cite{gemmeke2017audioset} as another baseline.
All models use a 4K-step linear warmup and linear decay to 40K steps; random codebook runs train for 40K steps while BEATs tokenizer runs converge by 20K. 
The learning rate is $5\times10^{-4}$, except for Table~\ref{tab:results} settings (6) and (7), which require a lower rate of $2.5\times10^{-4}$ for stable convergence.
Batches are sized to one H100 GPU: 25~min of audio for settings (2,\,3,\,5,\,7) and 50~min for (4) and (8).

\noindent \textbf{Evaluation:} To evaluate the learned representations, we use the Watkins marine mammal classification task over 31 species from the BEANS benchmark~\cite{hagiwara2023beans}. We adopt the standard BEANS train/validation/test split, selecting hyperparameters on the validation set and reporting accuracy on the test set.
We report accuracy under two evaluation protocols: 1) \textbf{Linear probing:} A linear classifier trained on frozen encoder embeddings from last layer. This directly measures representation quality \cite{noss, superb} learnt from pre-training. 
2) \textbf{Full fine-tuning:} The entire encoder is fine-tuned on the labeled training set.
\\
\noindent
\textbf{Results:} Results are presented in Table~\ref{tab:results}. 
\textbf{Saliency in data dominates scale.} Pre-training on the $30$\,hr of operator-verified Cut Tapes yields $72.0\%$ linear probing accuracy with the random codebook and $73.7\%$ with the BEATs tokenizer, against $43.1\%$ and $42.8\%$ for the $1{,}240$\,hr of Master Tapes. 
Restricting the corpus to confirmed vocalizations is worth roughly $29$ to $31$ points while using $41\times$ less audio, and correspondingly less pre-training compute.
Passive recordings contribute little over no pre-training at all. 
The Master Tapes lift linear probing from $33.6\%$ (no pre-training) to only $43.1\%$, a gain of $9.5$ points for $1{,}240$ hours of audio. 
The Cut Tapes deliver $38.4$ points from $30$ hours. 
Passive recordings are not merely less efficient per hour; the bulk of them contribute almost nothing.
In-domain passive audio is worse than out-of-domain audio. 
AudioSet, which contains no marine mammal recordings, reaches $65.5\%$ and $65.8\%$, over $22$ points above the in-domain Master Tapes. 
Yet the Cut Tapes surpass AudioSet ($72.0\%$ and $73.7\%$) with $180\times$ less data. 
Domain match is therefore valuable only when the data is salient: an in-domain corpus dominated by ambient noise is a worse pre-training signal than a large out-of-domain corpus of ordinary sound events. 
\textbf{A better tokenizer cannot compensate for non-salient data.} Replacing the random codebook with the BEATs-iter3 tokenizer leaves the Master Tapes representation unchanged ($43.1\%$ to $42.8\%$) while marginally improving the Cut Tapes ($72.0\%$ to $73.7\%$). 
Tokenizer quality and data quality do not compensate for one another: a tokenizer with learned acoustic structure has nothing to exploit when the frames it must describe are dominated by non-salient signal. 
We examine the mechanism behind this in Sec.~\ref{sec:rq4}.
Under full fine-tuning the gap between passive and salient pre-training narrows sharply, from $29$ points to $11.8$ with the random codebook ($79.4\%$ against $91.2\%$) and to $5.0$ with the BEATs tokenizer ($87.0\%$ against $92.0\%$). 
Supervised training therefore recovers much of what poor pre-training loses, but it does so by consuming labeled data, which is precisely the resource self-supervision exists to conserve.

\subsection{RQ2: Salient Event Detection in Passive Ocean Recordings}
\label{sec:rq2}

We benchmark the birdsong salient-event detectors of Sec.~\ref{sec:sed} by applying them to ocean bioacoustics to test whether they generalize.

\noindent \textbf{Setup:} We construct a strongly labeled ocean events retrieval dataset from the Master Tapes and Cut Tapes of Watkins.
Each Cut Tape is an operator-selected excerpt of a known call, but its timestamp within the Master Tape is not recorded, so we recover it with normalized cross-correlation (NCC) and manual verification.
Because a Master Tape's hydrophone channels are recorded simultaneously, a true match must occur at the same time on every channel. 
We leverage this and accept a localization only when at least two channels independently place the excerpt at the same offset above an NCC floor.
Following this procedure, we do not aim to recover every Cut Tape, but make sure the localizations we keep are highly reliable.
This yields $410$ localized events, each placed at a random position within a $10$\,s clip, paired with $410$ event-free negative clips drawn from the same recordings.
For reference we also evaluate on the seven strongly-annotated soundscape test sets distributed with BirdSet~\cite{birdset_iclr,hsn,nes,per,pow,sne,ssw,uhh}.
We benchmark four label-free detectors.\footnote{Benchmark code at \url{https://github.com/Shikhar-S/salient-ocean-events}.}
We run \texttt{MAAD-ROI} with \texttt{bambird}'s segmentation parameters~\cite{bambird} (preliminary experiments tuning did not meaningfully change performance).

\noindent \textbf{Evaluation:} A predicted event is matched to a ground-truth event when their temporal intersection-over-union (IoU) is at least $0.5$, with matching performed one-to-one and greedily in the order of descending score.
Sweeping the detection threshold traces a curve of \emph{recall}, the fraction of ground-truth events recovered, against \emph{selection rate}, the fraction of the audio timeline covered by predicted events. 
A good detector attains high recall at low selection rate, so we summarize each method by the area under this recall--selection curve (AUC). We use selection rate rather than precision or a standard receiver operating characteristic AUC (ROC-AUC) because our high-precision annotation does not label every salient event, making the true negative and false-positive counts those metrics require unavailable; timeline coverage, in contrast, is always measurable.
This AUC measure is inherently low in scale: with events this sparse, recovering more of them means flagging more of the timeline. MAAD-ROI (Table~\ref{tab:sed}), for instance, recovers only 40\% of events yet flags 69\% of the recording, so its curve stays shallow and its area is small.


\begin{table*}[t]
\centering
\caption{Salient event detection on passive ocean (Watkins) and bird (BirdSet) recordings. Events are matched at IoU $\geq 0.5$. \textbf{AUC} is the area under the recall-selection curve (higher is better); \emph{Recall} and \emph{Sel.} (selection rate) are read at the keep-all operating point. BirdSet columns are the mean over $7$
habitats. All methods are weak: the best detector recovers only $40\%$ of ocean events while flagging $69\%$ of the timeline. Please see Sec.~\ref{sec:rq2}.}
\label{tab:sed}
\small
\begin{tabular*}{\textwidth}{
@{\extracolsep{\fill}}
lcccccc
@{}
}

\toprule
& \multicolumn{3}{c}{\textbf{Watkins}}
& \multicolumn{3}{c}{\textbf{BirdSet}} \\
\cmidrule(lr){2-4}
\cmidrule(lr){5-7}

\textbf{Method}
& Recall $\uparrow$
& Sel. $\downarrow$
& \textbf{AUC} $\uparrow$
& Recall $\uparrow$
& Sel. $\downarrow$
& \textbf{AUC} $\uparrow$ \\

\midrule
\texttt{MAAD-ROI} \cite{ulloa2021scikit,maad_roi}         & 0.402 & 0.689 & \textbf{0.317} & 0.465 & 0.604 & \textbf{0.319} \\
\texttt{PCEN-Peak} \cite{pcen_peak}         & 0.288 & 0.753 & 0.209 & 0.253 & 0.600 & 0.181 \\
\texttt{Spectral Entropy} \cite{spectral_entropy}  & 0.224 & 0.398 & 0.205 & 0.100 & 0.313 & 0.089 \\
\texttt{Energy Template} \cite{energy_template}   & 0.259 & 0.618 & 0.152 & 0.161 & 0.588 & 0.098 \\
\bottomrule
\end{tabular*}
\end{table*}

\noindent \textbf{Results:} Table~\ref{tab:sed} shows the results.
\textbf{All methods leave substantial room for improvement.} The best detector, \texttt{MAAD-ROI}, reaches an AUC of only $0.317$ on Watkins.
Moreover, it only recovers $40\%$ of events by flagging $69\%$ of the timeline.
No method achieves high recall at low selection rate on either corpus.
The four detectors originate in four different domains, and the ordering on ocean acoustics follows none of them.
The best is \texttt{MAAD-ROI}, built for \emph{bird} activity detection ($0.317$); next is \texttt{PCEN-Peak}, adapted from far-field \emph{keyword spotting} ($0.209$); then \texttt{Spectral Entropy}, a \emph{soundscape ecology} index ($0.205$).
Last is \texttt{Energy Template} ($0.152$), which adapts the spectrogram-correlation principle of classical \emph{marine mammal} detection~\cite{energy_template} to the label-free setting.
This motivates treating salient event detection as a first-class problem in ocean acoustic SSL rather than an off-the-shelf preprocessing step.
Pre-training corpora are filtered by such detectors, and a filter that recovers $40\%$ of events while selecting $69\%$ of the timeline cannot isolate the signal a model should learn from.
We confirm this in Sec.~\ref{sec:rq3}, where representations learned from pre-training on \texttt{MAAD-ROI}-filtered data still trail pre-training on operator curated data by a wide margin.

\subsection{RQ3: Representation Quality using Filtered Data} \label{sec:rq3}
RQ2 showed that salient event detection methods localize events poorly.
Here we study whether ad hoc post-processing using these detectors can substitute for operator curation.
This also addresses an open question about whether passive recordings combined with saliency detection  filtering can serve as a viable pre-training corpus.

\noindent \textbf{Setup and Evaluation:} We filter the Watkins Master Tapes with \texttt{MAAD-ROI}, the strongest detector from RQ2. 
Each recording is tiled into non-overlapping $10$\,s windows and the detector is run independently on each. 
We score each window by the total power of its detected regions and keep the top $50\%$ within each recording. 
Ranking within a recording rather than globally prevents louder tapes from crowding out quieter ones. 
This reduces $1240$\,hr of Master Tapes to $613$\,hr.
The pre-training and evaluation setup is the same as RQ1.

\noindent \textbf{Results:} Table~\ref{tab:results} shows the results. 
\textbf{Saliency filtering slightly improves learned representations.} Saliency filtering discards half the audio, and raises linear probing from $43.1\%$ to $47.5\%$ and fine-tuning from $79.4\%$ to $90.6\%$. 
Selecting the more salient half of a passive recording is therefore better than using all of it.
\textbf{However, the quality does not approach human curation.} The Cut Tapes reach $72.0\%$ linear probing with only $30$\,hr, $24.5$ points above the filtered corpus while using $20\times$ less audio. 
Filtering closes only $4.4$ of the $28.9$-point gap between passive recordings and human curation. 
Under fine-tuning the filtered corpus reaches $90.6\%$, compared to $91.2\%$ from human annotation. 
The benefit of well-curated pre-training data thus appears in the representation, which is what self-supervision is meant to produce.
To summarize, ad hoc post-processing with weak salient event detection methods leaves a large room for improvement in representation quality.

\subsection{RQ4: Capacity Allocation to Non-Salient Audio}
\label{sec:rq4}

To understand how pre-training on non-salient data affects representations in the BEATs framework, we inspect the prediction targets that the SSL objective actually optimizes.
 
\noindent \textbf{Setup:} We tokenize the Master Tapes and the Cut Tapes with both tokenizers from Sec.~\ref{sec:rq1} and measure the target distribution over $4{,}000$ segments per corpus. 
Both tokenizers draw from a $1024$-entry codebook, bounding target entropy at $10$ bits. 
We compute and analyze the frame-marginal token entropy, the effective codebook size, the share of frames claimed by the five most frequent codes, and the rate at which consecutive frames repeat a code.

\noindent \textbf{Results:} We find that \textbf{background sounds do not collapse.}
The Master Tapes produce targets that are \emph{less repetitive} than the curated Cut Tapes ($16.9\%$ against $27.7\%$ of consecutive frames repeating a code) and \emph{less concentrated} (the five most frequent codes cover $53.8\%$ of passive frames against $66.5\%$ of salient frames).
\textbf{Encoder capacity is wasted on entropy that does not carry biological information.} 
The passive corpus produces \emph{higher} entropy targets than the salient corpus ($5.04$ against $4.32$ bits), yet the representations are $29$ points worse under linear probing ($43.1$ against $72.0$, Table~\ref{tab:results}).
Masked prediction allocates capacity in proportion to the acoustic variation it must predict, not in proportion to signal salience, so the encoder wastes capacity modeling the content irrelevant to the downstream task. 
A better tokenizer re-expresses the problem rather than removing it.
The BEATs tokenizer assigns the passive corpus $8.91$ bits against $9.03$ bits for the salient corpus.
Therefore, it describes ambient noise with almost the same richness with which it describes biological vocalizations.
Better tokenization raises the fidelity of the noise description alongside the signal description, without changing what fraction of capacity is aimed at background.
Since neither more data nor a better tokenizer redirects capacity toward the signals, the only effective lever is which segments enter the corpus at all. 
This is the capability detectors from Sec.~\ref{sec:rq2} lack, and hence the filtering in Sec.~\ref{sec:rq3} closes only part of the gap to operator curation.


\section{Conclusion}

Using BEATs as a testbed, we showed that self-supervised audio models in ocean acoustics suffer from \emph{noise overrepresentation}: when informative events are temporally sparse, the encoder spends most of its capacity modeling a background that is acoustically rich but biologically irrelevant.
Pre-training on 30 hours of curated vocalizations outperforms 1{,}240 hours of raw recordings by 29 points in linear probing.
Inspecting the tokens BEATs predicts explains why: the background noise is high-entropy, and masked prediction allocates capacity in proportion to acoustic variation rather than signal salience, so neither more data nor a better tokenizer redirects it toward the signal.
Improving the representation instead requires curation: selecting which segments enter the corpus.
This is what human annotation provides; however, existing salient event detectors we benchmarked on ocean audio are not yet strong enough to automate it, and detector-based filtering closes only part of the gap.
Because sparse events in long recordings also arise in other domains (e.g. industrial, seismic etc.), we argue that signal saliency needs to be treated as a core part of building these models. 

\section{Acknowledgment}
\label{sec:ack}
We appreciate the observations from Dr. Kyle Becker and the Multidisciplinary University Research Initiatives (MURI) Program for motivating this study.
This work used Delta and DeltaAI at NCSA through allocations CIS210014 and IRI120008P from the ACCESS program, supported by NSF grants \#2138259, \#2138286, \#2138307,
\#2137603, and \#2138296.

\bibliographystyle{IEEEtran}
\bibliography{references}

\end{document}